\documentclass[extra,onecolumn]{gji}
\usepackage[fleqn]{amsmath}
\usepackage{amsfonts,amssymb}
\usepackage{timet,epsfig,wrapfig,psfrag}
\usepackage{natbib}
\usepackage{symbol}
\usepackage{color}
\usepackage[normalem]{ulem}
\usepackage{booktabs}
\usepackage{multirow} 
\usepackage{pdfcolmk} 
\usepackage{hyperref}

\usepackage[paperheight=10in,paperwidth=8in,margin=1in,heightrounded]{geometry}

\usepackage{soul}

\newcommand{\w}{\mathbf}

\usepackage{gensymb}

\definecolor{dgreen}{RGB}{0,150,0}

\definecolor{dblue}{RGB}{0,0,200}

\definecolor{dkred}{RGB}{200,0,0}

\begin{document}

\bibliographystyle{chicago}


\title[A high-resolution discourse on seismic tomography]{A high-resolution discourse on seismic tomography}

\author[Andreas Fichtner, Jeroen Ritsema and Solvi Thrastarson]{
\parbox{\linewidth}{Andreas Fichtner$^1$, Jeroen Ritsema$^2$ and Solvi Thrastarson$^1$}\\
$^1$ Department of Earth and Planetary Sciences, ETH Zurich, Zurich, Switzerland\\
$^2$ Department of Earth and Environmental Sciences, University of Michigan, Ann Arbor, U.S.A.}

\maketitle


\begin{summary}
Advances in data acquisition and numerical wave simulation have improved tomographic imaging techniques and results, but non-experts may find it difficult to understand which model is best for their needs. This paper is intended for these users. We argue that our notion of best is influenced by the extent to which models satisfy our biases. We explain how the basic types of seismic waves see Earth structure, illustrate the essential strategy of seismic tomography, discuss advanced adaptations such as full-waveform inversion, and emphasize the artistic components of tomography. The compounding effect of a plethora of reasonable, yet subjective choices is a range of models that differ more than their individual uncertainty analyses may suggest. Perhaps counter-intuitively, we argue producing similar tomographic models should not be the goal of seismic tomography. Instead, we promote a \emph{Community Monte Carlo} effort to assemble a range of dissimilar models based on different modeling approaches and subjective choices, but which explain the seismic data. This effort could serve as input for geodynamic inferences with meaningful seismic uncertainties.
\end{summary}


\section{Introduction}\label{S:Introduction}

For many decades, global networks of seismographs have produced millions of seismograms rich in information on wave propagation through the Earth \citep{Butler_etal_2004,Leroy_2023,Wilson_2023}. Decoding this information at once into a model, or image of seismic velocity variations in the Earth, is a large research field known as \emph{seismic tomography}. Seismic tomography began in the 1970s \citep[e.g.,][]{Aki_Lee_1976,Dziewonski_1977,Aki_1977}. The first 3-D global images inspired "simple global models of plate dynamics and mantle convection" \citep[e.g.,][]{Hager_1981}, as well as estimates of viscosity, density, and chemical composition \citep[e.g.][]{Dziewonski_1977,Hager_1984}.
In the years before graphical user interfaces, seismologists applied rudimentary techniques to analyze seismograms, sometimes printed on paper and dissected using ruler and pencil. Those days are mostly behind us because the continuous influx of data from thousands of seismometers around the world requires automated data assimilation methods. As data sets have grown exponentially, tomographic images have increased in quality and number and motivated computer simulations to explain planetary-scale geology, the history of the motions of tectonic plates, and the composition and flow of rock in the deep interior of the Earth \citep[e.g.,][]{Bunge_2003,Bocher_2018,Ghelichkan_2021,Ghelichkan_2024}.

Numerous research articles and books are available on the technical aspects of seismic tomography and applications on local and global scales \citep[e.g.,][]{Iyer_1993,Nolet_2008,Fichtner_book}. This chapter is a modest attempt to capture the essence of seismic tomography and is intended for scientists who are keen on using the products of seismic tomography in research and teaching, but who are unfamiliar with the machinery and perhaps disoriented by the plethora of models and methods. For this, we use language which is as non-technical as possible to discuss the salient points of tomography. Our focus is on global-scale tomography, because it facilitates the presentation of the topic and the comparison of different tomographic model, which are all images of exactly the same object: the whole Earth. We do not cover tomography in the context of seismic exploration, which often benefits from ground-truth information in the form of well-ties. 

Section \ref{S:Data} aims to explain different types of seismic waves and the information they may carry about the internal structure of the Earth. Based on the example of travel time tomography, formulated by linear algebra, Section \ref{S:Methods} will introduce more recent concepts, such as finite-frequency tomography or full-waveform inversion, and the balancing act between resolution, effort, and computational cost. The diversity of data and methods translates into a wide range of tomographic models that agree in some respects, but disagree in others. Section \ref{S:Models} presents some of the most recent models and touches on important questions that users of tomographic models are likely to have. How good is a certain model? Which model is the best? Why are the models so different, and is that actually a problem?

Like many before us, we promote seismic tomography as a powerful approach to convert terabytes of data into images of the interior of the Earth. We have dabbled in the field of seismic tomography ourselves and recognize the potential for continuous improvements thanks to the expansion of networks (potentially into the oceans), new developments in theoretical and computational approaches, and ever increasing computational power. However, notwithstanding enthusiasm, we find it important to take stock of limitations that are difficult to overcome, and to draw a fine line between science and art. If we convey one takeaway message in this introduction, it is that the uncertainties in seismic tomography are significantly larger than estimated by individual practitioners based on statistical analyses, essentially owing to exactly the "artistic" component of seismic tomography. However, when properly understood and interpreted, these uncertainties may transform from problem to opportunity.

\section{A quiz}\label{S:Quiz}

\begin{figure}
\centerline{\includegraphics[width=1.0\textwidth]{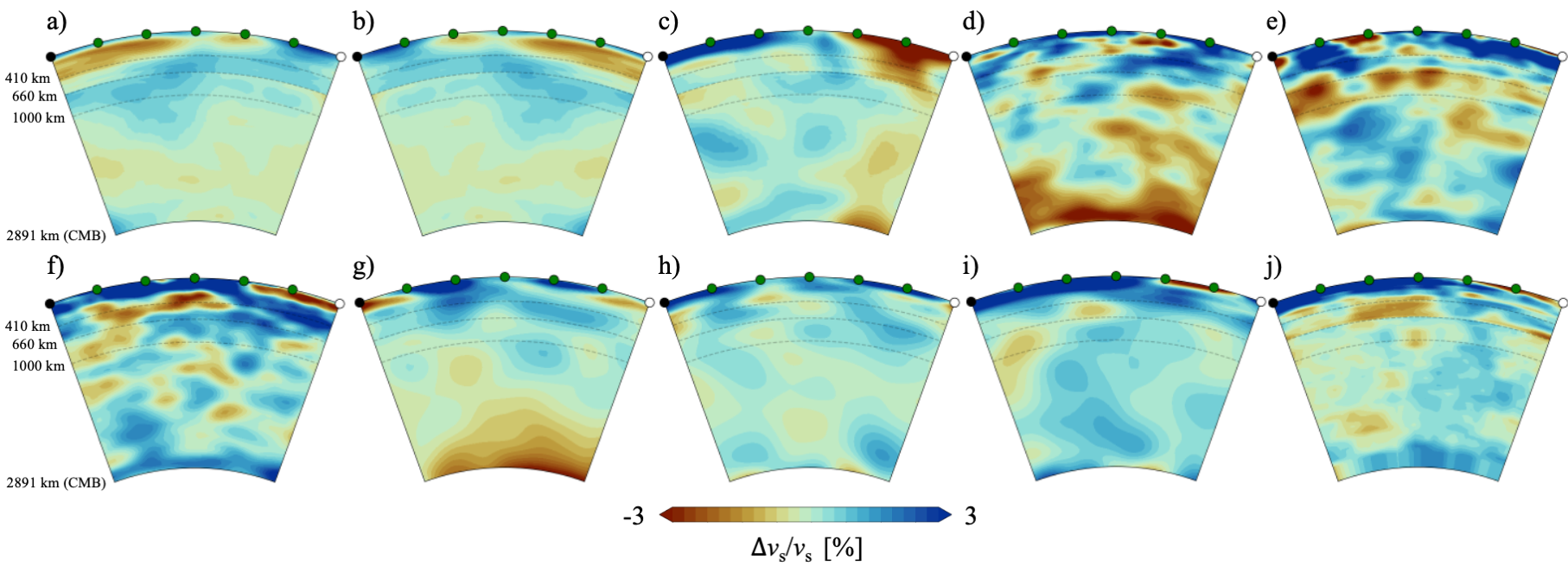}}
\caption{Collection of ten unlabeled vertical slices, $60^\circ$ wide and from the core-mantle boundary (CMB) to 50 km depth, through five recent global S velocity models \citep{Ritsema_2011,French_2014,Simmons_2021,Cui_2024,Thrastarson_2024}. Shown is the variation of S-wave velocity $\Delta v_\text{s}$ relative to the absolute S-wave velocity $v_\text{s}$ in the spherically symmetric Earth model PREM \citep{Dziewonski_Anderson_1981}. The reader is asked to identify the slices centered on an active subduction zone, where we expect a high-velocity descending slab.}
\label{fig:quiz_1}
\end{figure}

\begin{figure}
\centerline{\includegraphics[width=1.0\textwidth]{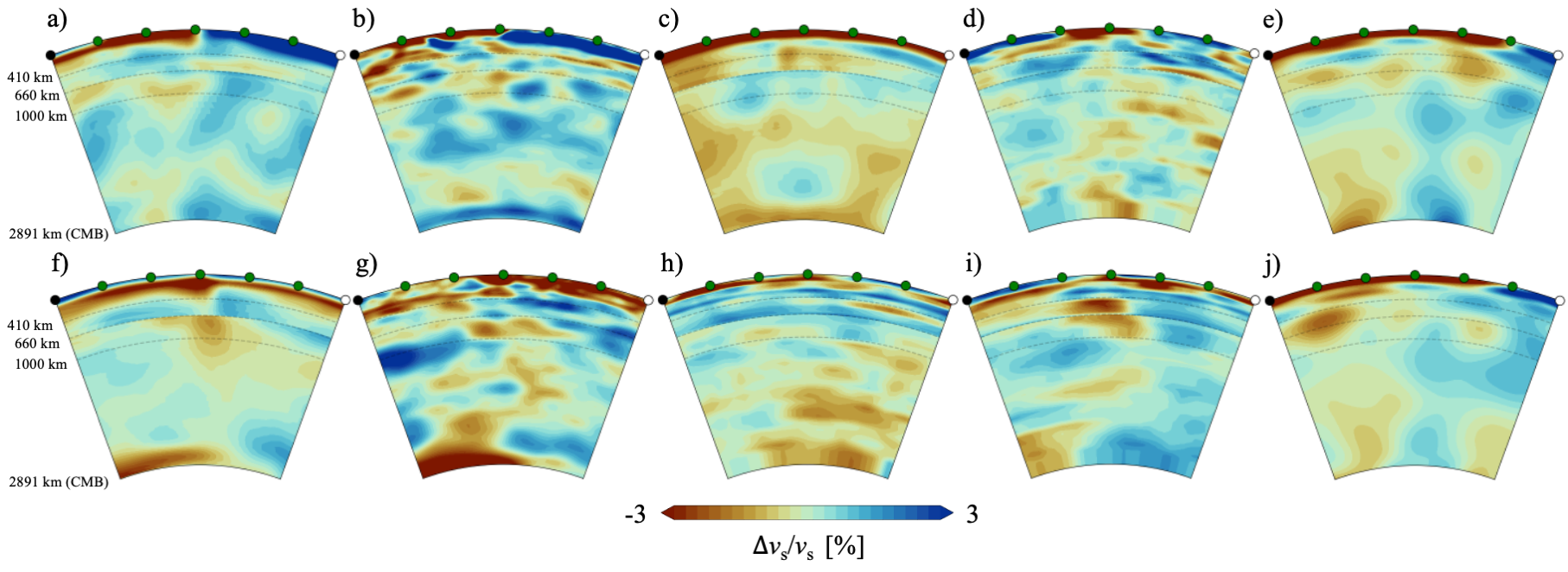}}
\caption{Collection of ten unlabeled vertical slices, $60^\circ$ wide and from the core-mantle boundary (CMB) to 50 km depth, through five recent global S velocity models \citep{Ritsema_2011,French_2014,Simmons_2021,Cui_2024,Thrastarson_2024}. Shown is the variation of S-wave velocity $\Delta v_\text{s}$ relative to the absolute S-wave velocity $v_\text{s}$ in the spherically symmetric Earth model PREM \citep{Dziewonski_Anderson_1981}. Which of these slices crosses an active hotspot thought to be caused by a mantle plume rising from the core-mantle boundary to the surface?}
\label{fig:quiz_2}
\end{figure}

We begin with a short quiz to step on some toes, including our own, and to let readers decide whether it is worth reading beyond this paragraph. Figs. \ref{fig:quiz_1} and \ref{fig:quiz_2} show vertical slices through five global tomographic models of the mantle \citep{Ritsema_2011,French_2014,Simmons_2021,Cui_2024,Thrastarson_2024} and address tomographic evidence for the descent of subducted oceanic lithosphere (i.e., "slabs") through the transition zone and the rise of plumes from the core-mantle boundary, topics frequently raised in the literature. Each correct answer to the following two questions is worth a point, so 20 points total can be earned. (i) Which of the sections in Fig. \ref{fig:quiz_1} crosses an active subduction zone? (ii) Which of the sections in Fig. \ref{fig:quiz_2} crosses an active hotspot? The solution to the quiz can be found in section \ref{SS:resolve_quiz}. Readers with 15 points or fewer ($\leq$ 75\%) may choose to continue reading.

The quiz illustrates why seismic tomography is not straightforward. The five models in Figs. \ref{fig:quiz_1} and \ref{fig:quiz_2} are based on seismograms from the global seismic network and are optimal explanations of the data based on the chosen methods. Nevertheless, the appearance of seismic structures is different even at large scales, which indicates that uncertainties are large. The quiz may also reveal that we are biased in determining what the mantle should look like. Most of us find it satisfying to recognize the canonical textbook structures of slabs and plumes in the seismic images, and it is human nature to deem models that confirm our biases to be the best.

\section{Seismic data}\label{S:Data}

\begin{figure}
\centerline{\includegraphics[width=1\textwidth]{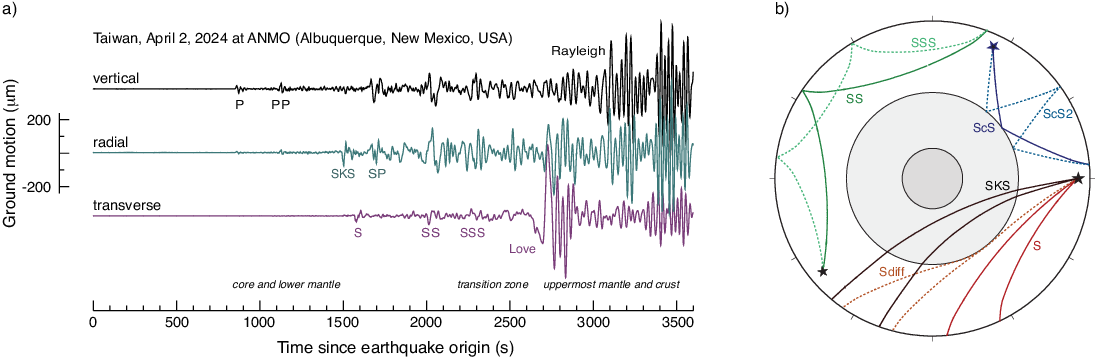}}
\caption{{\small{{
(a) Vertical, radial, and transverse ground displacement records of the April 2, 2024 Taiwan earthquake at seismic station ANMO (Albuquerque, New Mexico) at a distance of $106^\circ$. Several high-amplitude seismic signals are indicated in the hour-long record. (b) Ray paths of several shear waves, plotted for earthquakes at different locations, for better visibility. In the lower-right section of the globe are S (red), Sdiff, which is an S wave that has diffracted around the core (red, dashed), and SKS, which includes a P-wave segment in the outer core (black). In the upper-left section are SS (green) and SSS (green, dashed). In the upper-right section are ScS (blue) and ScS2 (blue, dashed).
}}}}
\label{fig:seismogram}
\end{figure}

The most important task in seismic tomography is, naturally, choosing the seismic data. In global-scale applications, seismologists choose recordings at epicentral distances larger than about $30^\circ$ ($1^\circ$ corresponds to about 111.2 km), the so-called \emph{teleseismic distances}, when the direct waves propagate through the lower mantle (e.g., S) or core (e.g., SKS). Usually, two processing steps are applied. First, the seismograms from the N-S and E-W channels of a seismometer are projected into the \emph{radial} (R) and \emph{transverse} (T) directions. This separates the \emph{P-SV} wave motions in the direction of the source from the \emph{SH} wave motions in the orthogonal direction. Second, the waveforms are filtered to suppress high-frequency signals related to earthquake rupture complexity and wave effects that cannot be explained by the \emph{initial model} (see Section \ref{S:Methods}).

Fig. \ref{fig:seismogram} is a typical example of a three-component teleseismic seismogram. It shows the vertical, radial, and transverse directions of the April 2, 2024 Taiwan earthquake recorded at seismic station ANMO (Albuquerque, New Mexico) at a distance of $106^{\circ}$. Two wave groups ''see'' different depth ranges of Earth's structure. The pulse-like signals between P and SSS are body waves that have propagated through the lower mantle and core. These body waves are followed by a train of higher-amplitude signals that form the surface waves. On the vertical-component (with up-and-down ground motions) and transverse-component (with sideways ground motions) seismograms, the surface wave is called the Rayleigh wave and Love wave, respectively.

\subsection{Surface waves}\label{S:Surface}

Seismic surface waves propagate horizontally through the crust and upper mantle akin to ripples along the water surface. Surface waves that have made multiple orbits around the globe can be visible after large earthquakes (magnitude $>$8), because they decay more slowly than body waves. The recorded surface wave train begins with slow and ends with fast oscillations. This phenomenon, called dispersion, is a distinctive characteristic of surface waves, caused by the strong wave speed increase with depth in the crust and upper mantle.

\begin{figure}
\centerline{\includegraphics[width=0.8\textwidth]{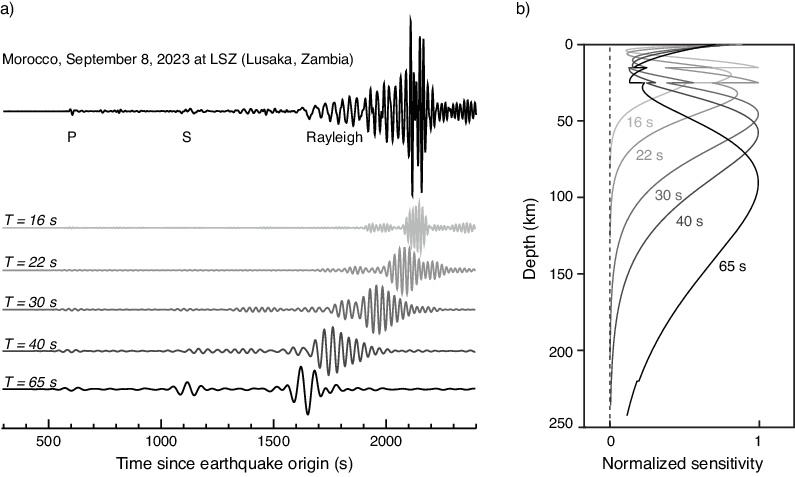}}
\caption{{\small{{(a) Vertical-component seismogram of the September 8, 2023 Morocco earthquake recorded at LSZ (Lusaka, Zambia) at an epicentral distance of $58\degree$. The upper trace shows the unfiltered waveform. The second, third, fourth, fifth, and sixth trace (from the top) are filtered waveforms with a dominant wave period $T$ indicated on the left. Note that low-frequency Rayleigh waves propagate faster than high-frequency Rayleigh waves. (b) Normalized sensitivity as a function of depth of the Rayleigh-wave phase velocity to shear-wave velocity for wave periods of 16 s, 22 s, 30 s, 40 s, and 65 s.
}}}}
\label{fig:dispersion}
\end{figure}

The record of the 8 September 2023 Morocco earthquake at seismic station LSZ in Zambia, shown in Fig. \ref{fig:dispersion}a, illustrates that long-period (i.e., low-frequency) surface waves propagate faster than short-period (i.e., high-frequency) surface waves. The Rayleigh wave through the African crust and upper mantle produces a long wave train with signals longer than 65 s period arriving around 1600 s after the earthquake and signals shorter than 20 s period arrive around 600 s later.

The sensitivity of the Rayleigh wave to wave speed structure depends on the wave period, as shown in Fig. \ref{fig:dispersion}b. Short-period surface waves ($<$20 s) propagate primarily through the crust. Surface waves with long periods ($>$60 s) penetrate deepest into the upper mantle, i.e., they have the largest \emph{skin depth}. This property can be used 
to estimate vertical variations of wave speed. 

\begin{figure}
\centerline{\includegraphics[width=1\textwidth]{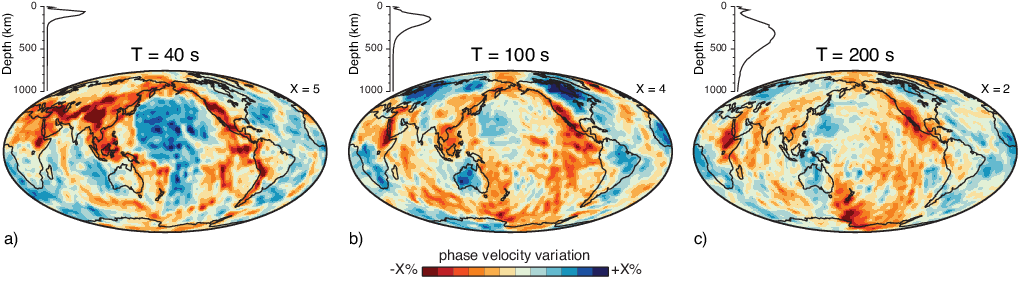}}
\caption{{\small{{Variation of the Rayleigh-wave phase velocity at periods of 40 s (a), 100 s (b), and 200 s (c). The normalized sensitivity of Rayleigh-wave velocity to shear-wave velocity in the upper 1000 km of the mantle is plotted on the top left.
}}}}
\label{fig:phasevel_maps}
\end{figure}

Fig. \ref{fig:phasevel_maps} shows maps of the Rayleigh wave speed at 40 s, 100 s, and 200 s period. The corresponding depth-dependent sensitivity is indicated above the maps. Rayleigh waves at 40 s period have a peak sensitivity around 60 km depth and their speed variations illuminate variations in, for example, the thermal structure of the oceanic lithosphere. A Rayleigh wave at a 100 s period has sensitivity to the shear wave structure in the upper 200 km of the mantle. Rayleigh wave speed variations at a period of 100 s expose, for example, the cold keels of the oldest crustal terranes. At 200-s period, Rayleigh waves see the structure below the tectonic plates. Low-wave-speed structures beneath eastern Africa, the Gulf of California, and the Southern Ocean south of New Zealand stand out and are related to dynamics deep in the upper mantle.

The finite and variable width of the Rayleigh wave sensitivity highlights limitations of surface wave data. Surface waves are sensitive to the crust at all periods, and the effect of the crust must be accounted for a priori (i.e., \emph{crustal corrections} must be applied) or, better still, wave speed variations in the crust and the depth of the crust-mantle boundary must be part of the tomographic model. The kernels are broadest for the longest periods, so vertical resolution decreases with depth. Overtone surface waves (i.e., most of the low-amplitude signals between the S wave and the large-amplitude fundamental-mode surface wave in Fig. \ref{fig:dispersion}a) help improve the vertical resolution, but the modeling of overtone surface waves is more difficult because they have similar arrival times and therefore their waveforms are intertwined. Hence, the quality of surface wave tomographic models between 300--700 km depth is not as high as in the upper 300 km of the mantle.

\subsection{Body waves}\label{S:Body}

Body waves at teleseismic distances constrain wave speeds in the lower mantle.
Body-wave signals appear as distinct wiggles, called \emph{body wave phases} before the surface waves. An earthquake produces a P wave and an S wave, but reflections and refractions off the surface and the core-mantle boundary lead to a sequence of body-wave phases identified by a sequence of capital letters not unlike genetic code. The P and S phases are direct paths from the source to the receiver. PP and SS have reflected once off the surface, and PPP and SSS have done so twice. PcP and ScS have reflected off the core and PKP and SKS have propagated through it. Not all phases are identifiable on a single seismogram, but a diverse collection of body-wave phases can be built using earthquakes with different focal depths and source mechanisms, and seismograms recorded over a wide epicentral distance range. Fig. \ref{fig:seismogram}b shows the ray paths of the most frequently used body waves in global S-wave tomography. S is the direct wave, including Sdiff that has diffracted along the core-mantle boundary, and SKS that propagated as a P wave through the outer core. SS and SSS are surface reflections, and ScS and ScSScS are reflections off the outer core boundary.

\begin{figure}
\centerline{\includegraphics[width=0.9\textwidth]{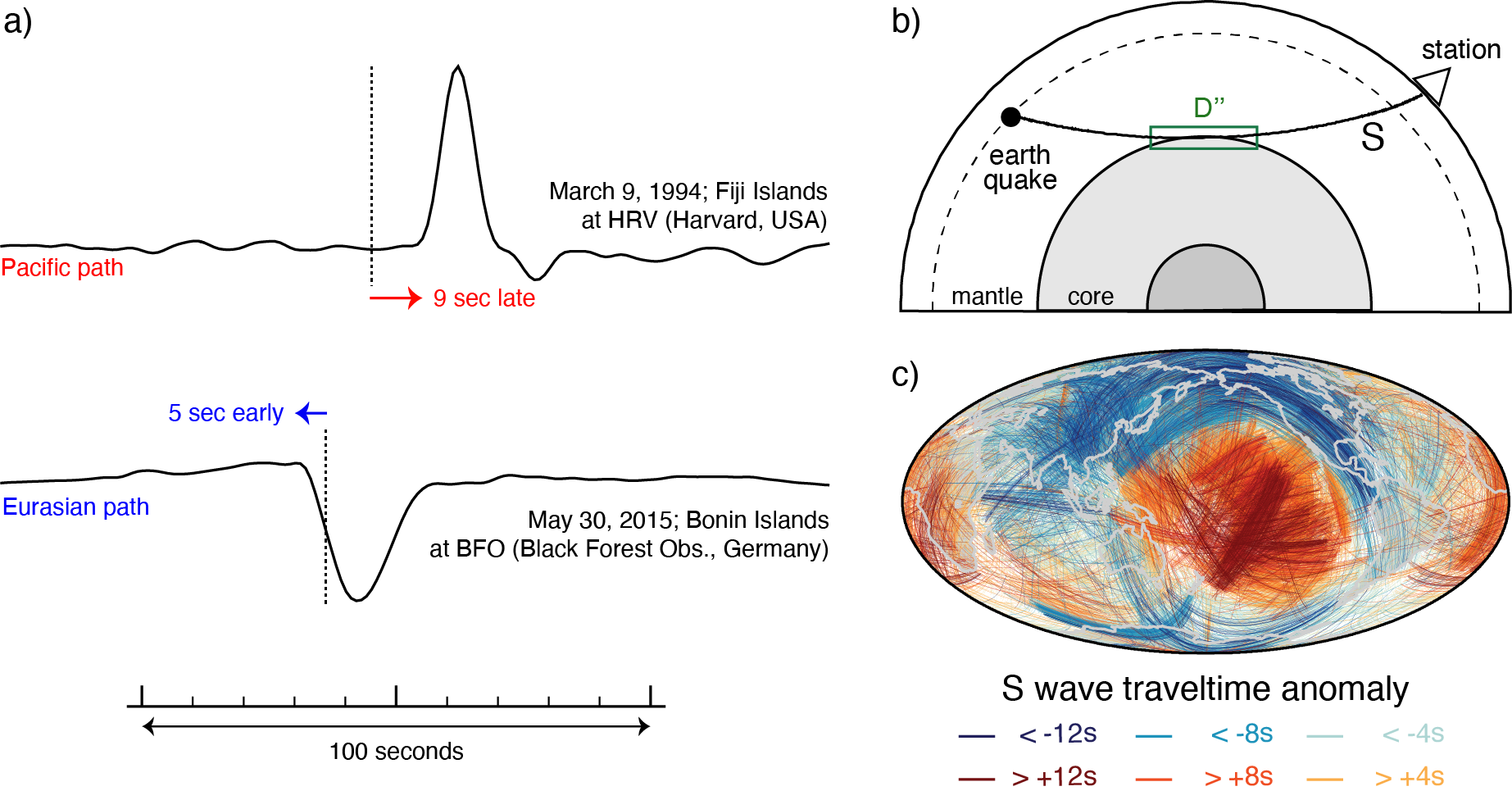}}
\caption{{\small{{(a) Transverse-component recordings of the diffracted S wave, Sdiff, generated by (top) the March 9, 1994 Fiji Islands earthquake recorded at HRV (Harvard, Massachussets, U.S.A.) and (bottom) the May 30, 2015 Bonin Islands earthquake recorded at BFO (Black Forest Observatory, Germany). The vertical lines are the predicted S wave arrival times for the PREM Earth model \citep{Dziewonski_Anderson_1981}. The horizontal arrows indicate the estimated delay times. The S waves propagate through the lower mantle beneath the Pacific and Eurasia, respectively. (b) Sketch of the S wave for an epicentral distance of $100^\circ$. The green box indicates the range of the S-wave segment in D'', the lowermost 300 km of the mantle. (c) Map of 35\,000 S-wave segments in the lowermost 300 km of the mantle from the S40RTS data collection \citep{Ritsema_2011}. The segments are colored blue and red if the S waves have shorter or longer travel times than predicted by PREM, respectively. The color intensity is proportional to the travel time anomaly.
}}}}
\label{fig:sdiff}
\end{figure}

Fig. \ref{fig:sdiff} illustrates the sensitivity of body waves to structure in the lower mantle. The isolated body-wave phase Sdiff is an S wave diffracted around the core. It propagates several hundreds of kilometers through the lowermost mantle, a region called D''. In the upper trace of Fig. \ref{fig:sdiff}a, Sdiff propagates through D'' beneath the Pacific Ocean about 9 s longer than predicted by the spherically symmetric Earth model PREM \citep{Dziewonski_Anderson_1981}. Sdiff for the trace below propagated through D'' beneath Eurasia and is recorded 5 s earlier than expected by PREM. Although wave propagation time delay can be accrued anywhere along the propagation path, it is likely that most of the delay originates in D'' because the map of 35\,000 Sdiff delays in Fig. \ref{fig:sdiff}c reveals a coherent pattern when the delays are plotted along the Sdiff segments in D''. Almost all Sdiff waves propagate slowly if they have propagated through D'' beneath Africa and the Pacific. These two structures, clearly visible without doing an inversion, are known as the \emph{large low shear-velocity provinces} (LLSVPs) of the lower mantle. Their role as large-scale mantle upwellings or as distinct geochemical reservoirs has been reviewed by \citet{McNamara_2019}.

The teleseismic body-wave phases along vastly different paths illuminate the lower mantle but there are significant limitations. Teleseismic body waves are affected by the upper mantle but cannot resolve it. All body waves propagate steeply through the upper mantle (incidence angles at the surface are smaller than $20\degree$). Body-wave rays in the upper mantle cross over only in regions where seismicity or station density is high. So, the depth extent of structures in the upper mantle cannot be constrained by teleseismic body waves. Teleseismic body-wave tomography has characteristic \emph{vertical smearing} in the upper mantle, which can easily be mistaken as the image of a mantle plume. In addition, large regions of the lower mantle are not sampled because earthquakes occur along plate boundaries, most seismic stations are on land, and Snell's law forces body waves to turn upwards away from the lower mantle. We will come back to the issue of body wave resolution in Section \ref{SS:model_quality}.

\section{Seismic tomography}\label{S:Methods}

Despite differences in modeling approaches and data types, the goal of seismic tomography is to construct  models of the wave speed in Earth's interior that explain the wiggles in seismograms, mostly the travel times of seismic waves. It is therefore not surprising that all tomographic methods share basic concepts. We will explain these in Section \ref{SS:basics} using the example of \emph{travel time tomography} where seismic wave propagation is approximated by \emph{ray theory}. More involved variants of seismic tomography, such as \emph{finite-frequency tomography} and \emph{full-waveform inversion}, are topics of Section \ref{SS:beyond_rays}. They mostly aim to improve the methods used to simulate the propagation of seismic waves and to quantify the relation between measurements and 3-D Earth structure. While the choice of seismic data and tomography method affect the resulting model the most, the combined effect of many seemingly minor adaptations can be substantial, as explained in Section \ref{SS:choices}.

\subsection{Methodological basics: Using the travel times of waves for imaging}\label{SS:basics}

According to ray theory, the travel time $t$ of a wave is the ratio of the wave propagation length $l$ and the wave velocity $v$ along an infinitely thin geometric ray between the wave source (e.g., the earthquake) and the wave recorder (i.e., the seismic station). Usually, we measure the difference, $\Delta t = t - t_0$, between the recorded travel time, $t$, and the travel time predicted for some initial or reference velocity model of the Earth, $t_0$. This difference, called a \emph{travel time anomaly}, can be related to wave velocity integrated along the ray path,
\begin{equation}\label{E:methods001}
\Delta t = t - t_0 = \int_\text{ray} \frac{dl}{v(l)} - \int_\text{ray} \frac{dl}{v_0(l)}\,,
\end{equation}
where $v$ and $v_0$ are the unknown velocity in the Earth and the velocity in the initial model, respectively. Travel time tomography requires a large number of travel time anomaly measurements to estimate the 3-D distribution of the wave velocity inside the Earth. In the simplest case considered here, this problem can be formulated in terms of linear equations. While this approach involves approximations that we will discuss in more detail in Section  \ref{SS:beyond_rays}, it allows us to solve the problem conveniently with the basic tools of linear algebra.

\subsubsection{Seismic tomography as a linear system of equations}

Eq. (\ref{E:methods001}) contains the reciprocals of velocity, which are difficult to handle mathematically. It is more convenient to work with the \emph{slowness}, $s=1/v$, and to rewrite Eq. (\ref{E:methods001}) as
\begin{equation}\label{E:methods002}
\Delta t = \int_\text{ray} s(l)\, dl - \int_\text{ray} s_0(l)\, dl = \int_\text{ray} [ s(l) - s_0(l)]\, dl = \int_\text{ray} \Delta s(l)\, dl\,. 
\end{equation}
Eq. (\ref{E:methods002}) describes the \emph{forward problem}. It predicts the travel time anomaly, $\Delta t$, for a given distribution of slowness anomalies, $\Delta s$, in the Earth. We want to, however, solve the \emph{inverse problem}, i.e., find the slowness anomaly $\Delta s$ from the travel time anomaly $\Delta t$. A common approach is to discretize the Earth into $M$ cells with a constant slowness and approximate the line integral in Eq. (\ref{E:methods002}) by a finite sum of $M$ terms
\begin{equation}\label{E:methods004}
\Delta t = \sum_{j=1}^M \Delta s_j \Delta l_{j}\,,
\end{equation}
where $\Delta s_j$ is the slowness anomaly and $\Delta l_{j}$ is the length of the ray segment in the $j^\text{th}$ cell. If the ray does not cross the $j^\text{th}$ cell, the length $\Delta l_{j} = 0$.

Fig. \ref{fig:basics}a shows an example for a simple model of the Earth comprised of
nine cells and a ray that traverses the $7^\text{th}$, $4^\text{th}$, $5^\text{th}$, $6^\text{th}$, and $3^\text{rd}$ cell between source and receiver. The travel time anomaly of the wave propagating along that ray is
\begin{equation}\label{E:methods005}
\Delta t = \Delta s_3 \Delta l_{3} + \Delta s_4 \Delta l_4 + \Delta s_5 \Delta l_5 + \Delta s_6\Delta l_6 + \Delta s_7 \Delta l_7 \,,
\end{equation}
arranged in order of cell index. The segment $\Delta l_4$ is the shortest, and $\Delta l_5$ the longest. It is common that a travel time data set includes thousands, if not millions, of measurements of $\Delta t$ for a wide range of source-station pairs and wave types. This means that we must solve equation (\ref{E:methods004}) many times.
\begin{equation}\label{E:methods005b}
\Delta t_1 = \sum_{j=1}^M \Delta s_{j} \Delta l_{1j}\,,\quad
\Delta t_2 = \sum_{j=1}^M \Delta s_{j} \Delta l_{2j}\,,\quad ...\quad
\Delta t_N = \sum_{j=1}^M \Delta s_{j} \Delta l_{Nj}\,.
\end{equation}
Mathematically, we write this long sequence of equations in index notation as
\begin{equation}\label{E:methods006}
\Delta t_i = \sum_{j=1}^M \Delta l_{ij} \Delta s_{j}\,,
\end{equation}
or as a \emph{matrix-vector} equation
\begin{equation}\label{E:methods007}
\w{d} = \w{L}\,\w{m}\,.
\end{equation}
The \emph{data vector}, $\w{d} = (\Delta t_1, \Delta t_2, \Delta t_3, \ldots, \Delta t_\text{N})$, includes the $N$ travel time measurements. The \emph{model vector}, $\w{m} = (\Delta s_1, \Delta s_2, \Delta s_3, \ldots, \Delta s_\text{M})$, includes the unknown slowness anomalies in the $M$ cells of the model. The \emph{sensitivity matrix} $\w{L}$ is a spreadsheet with $N$ rows and $M$ columns. The value of element $L_{ij}$ is the length of the ray segment through the $j^\text{th}$ cell for the $i^\text{th}$ travel time measurement.

\subsubsection{Solving the linear system, approximately at least}

We would like to directly invert $\w{L}$ to obtain $\w{m}=\w{L}^{-1}\w{d}$. However, it is impossible to find a model $\w{m}$ that matches $\w{d}$ exactly because measurements are imprecise and the chosen theory for wave propagation (e.g., ray theory) is an approximation. In fact, $\w{L}$ is not even a square matrix because the number of measurements, $N$, is typically much larger than the number of parameters, $M$, that describe the seismic structure of the Earth. 

To circumvent the inversion of $\w{L}$, we set a more modest goal. We estimate a model $\w{m}^\text{est}$ that predicts travel time anomalies $\w{d}^\text{est}=\w{L}\w{m}^\text{est}$ close to the observed ones, $\w{d}$. We define ``close to'' by the \emph{least-squares misfit function}
\begin{equation}\label{E:methods010}
\chi = \frac{1}{\sigma^2} \sum_{i=1}^N (d_i - d_i^\text{est})^2\,,
\end{equation}
which is the sum over squared differences between $N$ measured and computed travel time anomalies, $(d_i - d_i^\text{est})^2$, divided by the squared error, $\sigma^2$, due to noise or measurement uncertainties. Usually, the error is different for each measurement but we assume here that it is the same for all. Using Eq. (\ref{E:methods007}), we can rewrite Eq. (\ref{E:methods010}) in matrix-vector notation,
\begin{equation}\label{E:methods010b}
\chi(\w{m}) = \frac{1}{\sigma^2} [\w{d}-\w{L}\w{m}^\text{est}]^T [\w{d}-\w{L}\w{m}^\text{est}]\,.
\end{equation}
To find an $\w{m}^\text{est}$ that minimizes the misfit $\chi$, we differentiate Eq. (\ref{E:methods010b})
with respect to $\w{m}^\text{est}$ and
set ${\displaystyle \frac{d\chi}{d\w{m}} = 0}$ to get the \emph{least-squares solution}
\begin{equation}\label{E:methods011}
\w{m}^\text{est} = (\w{L}^T \w{L})^{-1} \w{L}^T \w{d}\,.
\end{equation}
The disappearance of $\sigma$ indicates that errors do not affect the estimated model. However, $\sigma$ will re-emerge when we discuss the quality of the estimated model in Section \ref{SS:model_quality}.

\begin{figure}
\centerline{\includegraphics[width=0.9\textwidth]{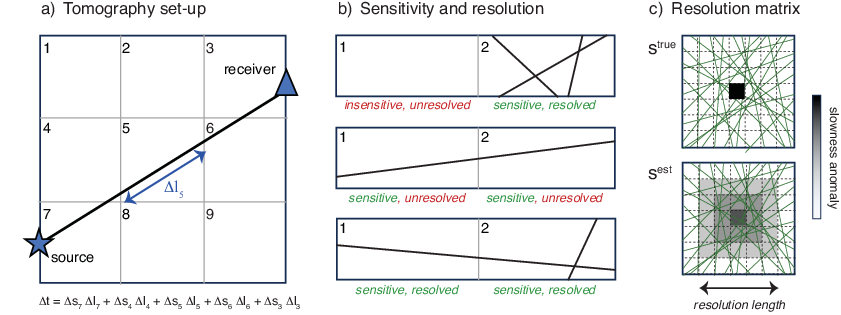}}
\caption{{\small{{Schematic illustration of tomography basics. a) Our domain of interest is discretized into 9 cells. The slowness anomaly $\Delta s_j$ within each cell $j$ is constant. A ray path (black line) connects a source and a receiver. The length of the ray segment within the $j^\text{th}$ cell is $l_{j}$, e.g., $\Delta l_5$ in the fifth cell. The forward modeling equation for the specific ray path shown here is displayed below. b) A simple Earth model consisting of two cells illustrates the difference between sensitivity and resolution. In the top panel, no ray crosses cell 1, meaning that there is neither sensitivity nor resolution. Multiple rays crossing cell 2 do not contribute information on cell 1, meaning that more data do not necessarily improve the results. In the middle panel, one ray path crosses cells 1 and 2, so the travel time is sensitive to the slowness in both cells. However, with only one measurement of $\Delta t$, we cannot resolve two unknown values $\Delta s_1$ and $\Delta s_2$ independently. In the lower panel, the two ray paths can be used to resolve both $\Delta s_1$ and $\Delta s_2$. c) The resolution matrix $\w{R}$ maps a hypothetical true model $\w{s}^\text{true}$ (top) into an estimated model $\w{s}^\text{est}$ (bottom). If the true model is a single-cell perturbation, as in the case shown here, the estimated model is called the \emph{point-spread function} and the width of the point-spread function is called the \emph{resolution length}.}}}}
\label{fig:basics}
\end{figure}

In contrast to $\w{L}$, the matrix $\w{L}^T\w{L}$ is square. Its inverse may exist in theory but, in practice, it does not because the rays cross the cells unevenly.
Consider, for example, the top panel of Fig. \ref{fig:basics}b where not a single ray traverses cell number 1, so the slowness $m_1$ is unconstrained by any observations. Saying the same in more technical terms, the data are exactly insensitive to $m_1$, and as a consequence $m_1$ is \emph{unresolved}. While sensitivity is necessary for resolution, it is not sufficient. This is illustrated in the middle panel of Fig. \ref{fig:basics}b. If two neighboring cells, e.g., with indices $j=1$ and $j=2$, are traversed by one and the same ray, the travel time of that ray is sensitive to both $m_1$ and $m_2$, but $m_1$ and $m_2$ cannot be resolved independently. If, for example, $\Delta t = 2$ and $l_1 = l_2 = 1$ the equation to solve is $2 = \Delta s_1 + \Delta s_2$, with infinite possible combinations for $\Delta s_1$ and $\Delta s_2$.

To make $\w{L}^T\w{L}$ invertible, it is common to add a scaled version of the unit matrix $\w{I}$ to $\w{L}^T\w{L}$ in Eq. (\ref{E:methods011}) and to approximate the least-squares solution as
\begin{equation}\label{E:methods012}
\w{m}^\text{est} = (\w{L}^T \w{L} + \varepsilon \w{I})^{-1} \w{L}^T \w{d} = \tilde{\w{L}}^{-1} \w{d}\,,
\end{equation}
where the \emph{damping factor} $\varepsilon$ is a small positive number and the \emph{generalized inverse} is $\tilde{\w{L}}^{-1} =(\w{L}^T \w{L} + \varepsilon \w{I})^{-1} \w{L}^T $.
This is equivalent to adding the term $\varepsilon (\w{m}^\text{est})^2$ to Eq. (\ref{E:methods010}),
\begin{equation}\label{E:methods013}
\chi(\w{m}) = \frac{1}{\sigma^2} (\w{d} - \w{d}^\text{est})^2 + \varepsilon (\w{m}^\text{est})^2\,,
\end{equation}
which means that models with large slowness anomalies, i.e., models that wander too far from the initial model,
are deemed undesirable. Returning to the middle panel of Fig. \ref{fig:basics}b, the optimal solution to $2 = \Delta s_1 + \Delta s_2$ is $\Delta s_1 = 1$ and $\Delta s_2 = 1$, because the vector $\w{m} = (1, 1)$ is the shortest vector that matches the data.  

Damping is one of many mathematical forms of \emph{regularization}. Another one is \emph{smoothing}, which forces the slowness values to be similar in adjacent cells. In the top panel of Fig. \ref{fig:basics}b, the slowness anomalies $\Delta s_1$ and $\Delta s_2$ would be the same if the data were inverted under a smoothness constraint. Any form of regularization is in essence subjective. It may seem reasonable to force a model to be small or smooth, but these impositions are not informed by seismic data or geological observations and may introduce unrecognizable imaging artefacts.

\subsubsection{How good is the model? - Data fit and the resolution matrix}\label{SS:model_quality}

Any estimated model $\w{m}^\text{est}$, no matter how it was constructed, raises an obvious question: How good is it actually? The most important metric for the quality of a model $\w{m}^\text{est}$ is the reduction of misfit $\chi$ that it achieves. Ideally, the difference in travel time anomalies, $d_i - d_i^\text{est}$, should approximately equal the measurement errors, $\sigma$, at least on average over all measurements $i$. For example, if the measurement errors are around 0.1 s, it is pointless to construct a model that explains the travel time anomalies to within 0.01 s. The model would \emph{over-fit} the data. Conversely, if $d_i - d_i^\text{est} = 1$ on average,
the model would $\emph{under-fit}$ the data. Ideally, $d_i - d_i^\text{est} \approx \sigma$, so 
\begin{equation}\label{E:methods014}
\chi \approx \frac{1}{\sigma^2} \sum_{i=1}^N \sigma^2 = N\,.
\end{equation}
Hence, the regularisation parameter $\varepsilon$ should be tuned such that model $\w{m}^\text{est}$ produces a least-squares misfit close to $N$.

The \emph{resolution} quantifies how well $\w{m}^\text{est}$ represents actual slowness structures $\w{m}^\text{true}$ in the Earth. In resolution tests, we assume a test structure that produced our observations $\w{d}$ via $\w{d}=\w{L}\w{m}^\text{true}$. Substituting $\w{d}$ into the estimation (\ref{E:methods012}) yields
\begin{equation}\label{E:methods013}
\w{m}^\text{est} = \tilde{\w{L}}^{-1}\, \w{L} \w{m}^\text{true} = \w{R}\, \w{m}^\text{true} \,.
\end{equation}
The \emph{resolution matrix} $\w{R}$ connects the true slowness model $\w{m}^\text{true}$ to the model that we estimate, $\w{m}^\text{est}$. It allows us to ascertain how a hypothetical model of the Earth projects into a blurry tomographic image of the Earth. Frequently, seismologists choose a checkerboard of alternating low and high slowness values or they imagine a structure reminiscent of geology. In a \emph{spike test} where the slowness anomaly differs from zero in a single cell only, the resulting image is known as the \emph{point-spread function}, which has a characteristic \emph{resolution length}, as illustrated in Fig. \ref{fig:basics}c.

\subsection{Beyond straight rays}\label{SS:beyond_rays}

While Section \ref{SS:basics} covers the basic concepts relevant for seismic tomography, the complexity of both seismic data and 3-D wave propagation demands methodological refinements of seismic tomography. Most of these improve the sensitivity matrix $\w{L}$ to more accurately incorporate the physics of wave propagation.

\subsubsection{Curved ray paths and iterative model improvements}

Ray paths are not straight, as assumed in Fig. \ref{fig:basics}. Rather, rays in the Earth are curved because the wave speed changes with depth. Fig. \ref{fig:seismogram}b illustrates the bent body-wave paths for the spherically symmetric Earth model PREM \citep{Dziewonski_Anderson_1981}. There is a complication, albeit one of second-order importance. The actual ray paths may be different from those computed for a (1-D) initial model. Therefore, the two integrals in Eq. \ref{E:methods001} are over different paths $s$. Ray-path perturbations have the effect of placing slowness anomalies in the wrong places. Iteration by stepwise improving both the model and the ray paths is a common solution
\citep[e.g.,][]{Bijwaard_2000,Widiyantoro_2000,Gorbatov_2001}. Starting with a hopefully accurate initial model, typically a 1-D model like PREM, we can compute initial ray paths and obtain a first estimate of 3-D Earth structure by approximately solving the linear system, as in (\ref{E:methods012}). We use this first estimate to recompute the ray paths and the corresponding sensitivity matrix $\w{L}$, obtain an improved model, and iterate until we explain the data to within their errors.

\subsubsection{Finite-frequency tomography}

The approximation of a wave by a ray is borrowed from optics, where it works well because the wavelength of high-frequency electromagnetic waves (several hundred nanometers for visible light) is many orders of magnitude smaller than the propagation distance. Seismograms, on the other hand, are rich in low frequencies. Therefore, seismic wavelengths are often only 10 - 100 times smaller than the distance between source and receiver. In this case, a whole range of phenomena, collectively referred to as \emph{finite-frequency effects}, come into play. The two most important ones are \emph{wave front healing} and \emph{spatially extended sensitivity}.

\begin{figure}
\centerline{\includegraphics[width=1.0\textwidth]{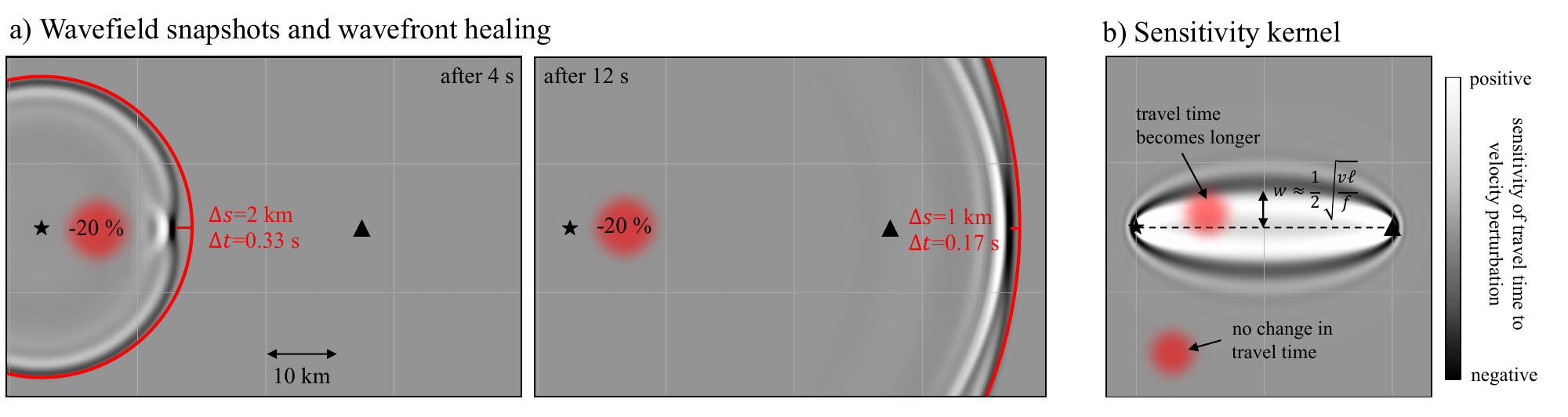}}
\caption{Illustration of finite-frequency effects using a numerical simulation of a wave (amplitudes in gray scale) with a dominant frequency of 2 Hz and a background velocity of 6 km$/$s. The source is indicated by a black star, and the receiver by a black triangle. a) Wavefront healing: A -20 \% velocity anomaly, indicated by the red blob, leads to a travel time delay that 'heals' over time. After 4 s (left panel), the wavefront is indented by 2 km, relative to the reference wavefront for a purely homogeneous medium, marked by the red circle. This corresponds to a travel time anomaly of $\Delta t=0.33$ s. After 12 s (right panel) the wave front has healed significantly. It is now indented by only 1 km, corresponding to a 0.17 s travel time anomaly. b) 'Fat ray' sensitivity: Instead of being sensitive to velocity perturbations along the infinitely thin ray path (dashed line), the travel time is sensitive to velocity perturbations within an extended sensitivity kernel. A negative velocity perturbation outside the kernel has no effect on the travel time. When located inside the kernel, it reduces the travel time of the wave. The width of the kernel is approximately given by $w\approx\sqrt{vl/f}/2$, where $f$ is the frequency of the wave.}
\label{fig:finite_frequency}
\end{figure}

Wavefront healing is illustrated in Fig. \ref{fig:finite_frequency}a, for a wave propagating through a low-velocity anomaly. Immediately after traversing the anomaly, the wavefront is indented by 2 km, corresponding to a travel time delay of 0.33 s. According to ray theory, this delay should be preserved, but the wavefront actually 'heals' as the wave continues to propagate. At the receiver, the travel time delay has reduced to 0.17 s. Finite-frequency methods for seismic tomography account for this effect \citep[e.g.,][]{Tong_1998,Dahlen_2000,Hung_2001}
and, with all else equal, the slowness anomalies in finite-frequency models are stronger than in ray-theoretical models \citep[e.g.,][]{Montelli_2004b}.

The wavelength of a wave, i.e., the space that it occupies, is $\lambda=v/f$, where $f$ is frequency. Consequently, the volume where a wave senses the properties of the medium increases with decreasing frequency. This volume, referred to as a \emph{sensitivity kernel} or \emph{influence zone} \citep[e.g.,][]{Yomogida_1992,Yoshizawa_2005}, is similar to the \emph{Fresnel zone} in optics, which is the region where obstacles affect the transmitted signal. For a homogeneous medium, the width of the Fresnel zone can be approximated as $w\approx\frac{1}{2}\sqrt{\frac{vl}{f}}$. In the example of Fig. \ref{fig:finite_frequency}b, we have $w\approx 7$ km. When the medium is more complicated, like the Earth, the exact sensitivity kernel can be computed using a mathematical trick known as the \emph{adjoint method} \citep[e.g.,][]{Tromp_2005,Plessix_2006,Fichtner_2006a}. Although these sensitivity kernels may have a complex shape, their width may still be estimated as $w\approx\frac{1}{2}\sqrt{\frac{vl}{f}}$, with $v$ some average velocity along the source-receiver path of length $l$. For example, a 0.1 Hz wave that travels 10$\,$000 km has a Fresnel zone that is several hundred kilometers wide. While sensitivity kernels represent the physics of seismic wave propagation more accurately than rays, the benefits for tomographic resolution are likely application-specific and somewhat debated \citep[e.g.,][]{vanderHilst_2005,Trampert_2006}.

\subsubsection{Full-waveform inversion}

\emph{Full-waveform inversion} (FWI) refers to two ideals that may not be reached very often in practice: The accurate simulation of the complete seismic wavefield in a 3-D heterogeneous Earth model with errors that are negligible compared to the data noise, and the exploitation of complete seismograms, i.e., the use of all information available in seismic recordings. Although the concept of FWI has been proposed already in the late 1970s and early 1980s \citep[e.g.,][]{Bamberger_1977,Bamberger_1982}, applications to 3-D seismic tomography had to await the advent of supercomputers powerful enough to enable numerical simulations of seismic wave propagation at reasonably high frequencies, i.e., $f>0.1$ Hz for regional and $f>0.01$ Hz for global applications, as order of magnitude \citep[e.g.,][]{Chen_2007,Fichtner_2009a,Tape_2010,French_2013}. This is because the computational requirements of a wavefield simulation scale as $f^4$. Mathematically, this unfavorable scaling originates from the \emph{Courant-Friedrichs-Lewy (CFL)} condition, which controls the stability of numerical simulations \citep[e.g.,][]{Quarteroni_2000,Fichtner_book,Igel_2016}, but it can also be understood intuitively. Increasing frequency by a factor of 2, reduces the spatial wavelength and the temporal period by a factor of 2. Hence, we require $2\times 2\times 2$ times as many grid points for the three space dimensions, and times steps that are only half as long. To obtain a seismogram of a given length, we therefore need $2^4=16$ as many mathematical operations.

FWI is typically implemented as an iterative process. It begins with the computation of synthetic seismograms for a plausible initial Earth model and their comparison to observed waveforms using a misfit function that measures the difference between the two time series. Sensitivity kernels computed with the adjoint method define the regions in the Earth where the misfit is sensitive to Earth structure, i.e., where changes in seismic velocity reduce the misfit. Following a first update of the Earth model, synthetic seismograms and sensitivity kernels are recomputed, and the procedure is repeated until the data fit reaches an acceptable value. Hence, FWI can be understood as a finite-frequency tomography that exploits more than selected wave arrivals and is improved iteratively.

As in the case of iterative travel time and finite-frequency tomography, the benefits of FWI are problem-dependent. The advantages of FWI become significant in the presence of sharp contrasts and velocity variations typically exceeding $\sim$10 \%, so that ray theory fails to accurately simulate seismic wave propagation. This is mostly the case in the lithosphere. Below the lithosphere, the amount and careful selection of data tend to be more important than the choice of one or the other tomographic method.

\subsection{All the little choices}\label{SS:choices}

Our simplified description of tomographic methods hides many of the little subjective choices that need to be made in practice. While the effect of each individual one may be deemed small, the fact that differences between models are often larger than suggested by formal resolution analyses, indicates that their combined impact may actually dominate model uncertainties. In the following we provide an unavoidably incomplete list of technicalities that enter the construction of any tomographic model.

Tomographic inverse problems are generally ill-posed, meaning that infinitely many models explain the data to within their errors. Hence, inversion algorithms need to be regularized to ensure convergence towards a meaningful model. Defining what is a meaningful model is inherently subjective, and there is no 'correct' way of implementing regularization. It is guided by prior expectations and pragmatism. Common forms of regularization include damping towards a preferred model, smoothing, and the omission of model parameters that are deemed unresolvable or not interesting. 

In- or excluding certain model parameters encodes subjective prior expectations about potentially interesting discoveries. For example, when we are interested in 3-D attenuation structure, we may disregard variations in density and several parameters describing azimuthal anisotropy, because it is unrealistic to resolve all of them at the same time. Although each of these parameters may have a small influence on the data, their combined effect may not be negligible \citep{Fichtner_2024}. 

The use of digital computers and the finite amount of data that we have available force us to discretize the Earth model. Similar to regularization, discretization is subjective, and it dictates the features that can or cannot be represented. A prominent example is the discretization of PREM \citep{Dziewonski_Anderson_1981} in terms of low-order polynomials that does not permit the representation of non-polynomial 1-D Earth structure that available data may request \citep{Kennett_1995}. For 3-D Earth models, discretizations in terms of spherical harmonics \citep[e.g.,][]{Ritsema_1999} and spherical splines \citep[e.g.,][]{Wang_1995} are widely used. Models constructed with FWI often co-use the numerical simulation mesh for the representation of Earth models to reduce workflow complexity \citep[e.g.,][]{Lei_2020,Thrastarson_2024}.

At global scale, the Earth's crust is too thin to be resolved and too thick to be ignored in seismic wavefield simulations. Correctly modelling the effect of the crust with ray theory is difficult because it contains strong small-scale heterogeneities. In numerical simulations, properly accounting for the crust requires a large number of closely spaced grid points, which increases computational cost. Consequently, seismograms are often corrected for the effect of the crust \citep[e.g.,][]{Bozdag_2008,Lekic_2010} or a smoothed version of the crust is implemented in numerical simulations \citep[e.g.,][]{Fichtner_2008,French_2014}. All of these strategies introduce errors, and the selection of one of them is again somewhat subjective.

Our choice of the least-squares misfit function in section \ref{SS:basics} was mostly dictated by convenience. It led to a linear system of equations that we know how to solve approximately. Other misfit functions, involving, for example, the amplitudes of seismic waves or the absolute values of travel time differences instead of their squares, would have been legitimate choices, too. The construction of misfit functions that find a useful balance between convenience and the extraction of robust information from seismic waveforms received considerable attention in the development of FWI methods. The number of options is large \citep[e.g.,][]{Gee_1992,Fichtner_et_al_2008,Bozdag_2011,Metivier_2016}, and each of them produces slightly different results.

\section{Earth model examples}\label{S:Models}

Seismic tomography is similar to painting a picture of a still life.
The paint, the type of brush, and the style (the tomographic method) determine the character of the painting even before the first brushstroke is on the canvas.
The choices are not entirely arbitrary, but are constrained by the artist's resources and time. There is no point in using the most sophisticated technique when the artist can only afford the cheapest materials (noisy data). Regardless of these choices, the artwork will always be artwork and not reality, unless the artist manages to paint at subatomic scale. Subjectively, we may prefer one painting to another. Objectively, however, the painting of a minimalist who focuses on the essence of few features is as prizeworthy as the painting of a realist who draws objects that may actually not exist.

\begin{figure}
\centerline{\includegraphics[width=1.0\textwidth]{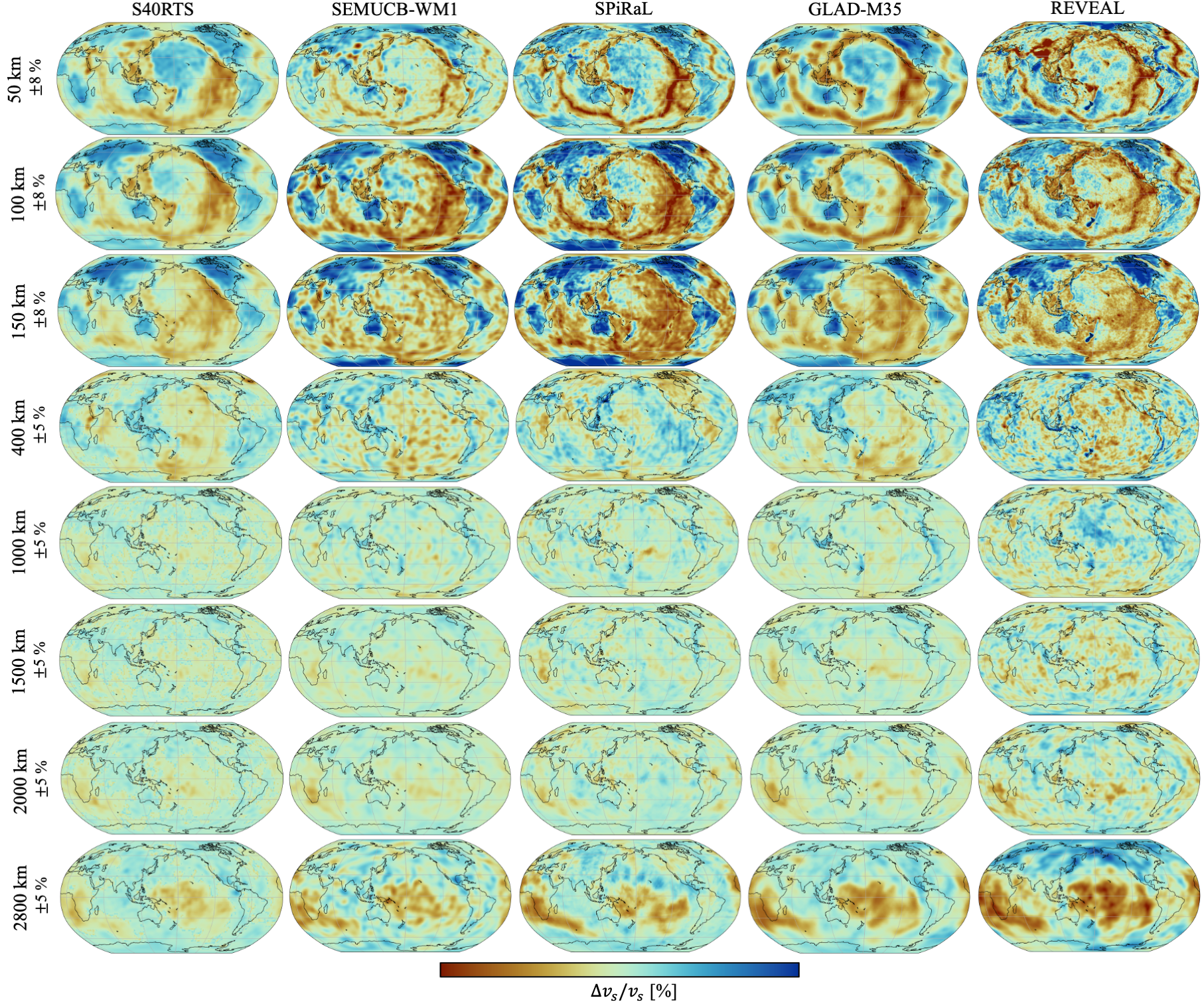}}
\caption{Horizontal slices through five recent global S velocity models, S40RTS \citep{Ritsema_2011}, SEMUCB-WM1 \citep{French_2014}, SPiRaL \citep{Simmons_2021}, GLAD-M35 \citep{Cui_2024} and REVEAL \citep{Thrastarson_2024}.}
\label{fig:model_collection}
\end{figure}

Fig. \ref{fig:model_collection} presents canvases from five different research groups. Each column shows maps of the variation of S velocity in the Earth's mantle, each based on a specific data collection, and choices in model construction; S40RTS \citep{Ritsema_2011}, SEMUCB-WM1 \citep{French_2014}, SPiRaL \citep{Simmons_2021}, GLAD-M35 \citep{Cui_2024} and REVEAL \citep{Thrastarson_2024}. Table \ref{tab:models} provides technical information about these models, including the type and amount of data, the tomographic method, and the parameterization used to construct these models. 
%
\begin{table}
    \centering
    \begin{tabular}{ |p{2cm}|p{6cm}|p{3cm}|p{2cm}| }
    \hline
        \textbf{model} & \textbf{type and number of data} & \textbf{tomographic method} & \textbf{model parameters} \\
        \hline\hline
        S40RTS & Rayleigh wave dispersion (40 - 320 s period, $\sim$26 million), S wave traveltimes ($\sim$400\,000), normal-mode splitting functions & travel time ray tomography combined with normal-mode inversion & S velocity \\
        \hline
        SEMUCB-WM1 & traveltimes of body waves (32 - 300 s period) and surface waves (60 - 400 s period) from 273 earthquakes and 500 stations & FWI & S velocity, radial anisotropy\\
        \hline
        SPiRaL & $\sim$4 million P wave and $\sim$50\,000 S wave travel times, surface wave dispersion maps & travel time ray tomography & P and S velocity, radial anisotropy \\
        \hline
        GLAD-M35 & frequency-dependent travel times of P, S and surface waves (17 - 250 s period) from 2\,160 earthquakes and 11\,575 stations & FWI & P and S velocity, radial anisotropy \\
        \hline
        REVEAL & time- and frequency dependent phase shifts of complete seismograms (30 - 100 s period) from 2\,366 earthquakes and 27\,879 stations & FWI & P and S velocity, radial anisotropy \\
        \hline
    \end{tabular}
    \caption{Summary of type and number of data, tomographic method and parametrization used in the construction of S40RTS \citep{Ritsema_2011}, SEMUCB-WM1 \citep{French_2014}, SPiRaL \citep{Simmons_2021}, GLAD-M35 \citep{Cui_2024} and REVEAL \citep{Thrastarson_2024}.}
    \label{tab:models}
\end{table}
%
It is obvious that the models portray the same planet. Plate tectonics and the ocean/continent dichotomy determine the seismic structure of the upper 400 km of the mantle. There are two antipodal "large-low velocity provinces" (or LLSVPs) at the base of the mantle. The mantle between 1000 km to 2000 km depth is the "boring middle" where the velocity variations are much smaller than in the uppermost and lowermost mantle. These characteristics are shared by virtually all S-wave models of the mantle, not just the five models displayed in Fig. \ref{fig:model_collection}.

However, the magnitude and sharpness of the velocity anomalies are different, and some models include intricate structures that are absent in others. Possibly, one model paints certain regions of the mantle more realistically than others, but we have to consider \emph{all the little choices}. How are models parameterized and discretized? How are the models regularized? How is anisotropy implemented and how are elastic parameters that are not included in the parameterization (e.g., density, attenuation) scaled to other parameters? Are the effects of the crust removed a priori by a global crustal model, or is it inverted for? What is the starting model and does it include inaccuracies (e.g., discontinuities or anomalous gradients) that still determine the model?

For example, S40RTS discretizes vertical velocity variations with splines that broaden with increasing depth. This may explain why the LLSVPs are weakest in S40RTS among the five models.  SEMUCB-WM1 and REVEAL implement a smoothed initial version of the crust, leaving it to the data to control the sharpness of the crust-mantle transition during the inversion process. In contrast, GLAD-M35 implements an actual discontinuity between the crust and the mantle, which is numerically more accurate but does not allow the data to easily adjust crustal thickness. Different treatments of the crust contribute to model differences mostly within the top 100 km that are primarily constrained by surface waves, which are particularly sensitive to crustal structure.

All sections in Figs. \ref{fig:quiz_1} and \ref{fig:quiz_2} show characteristics that we suspect are unconstrained by seismic data. For example, Fig. \ref{fig:quiz_2}d includes alternating high- and low-velocity anomalies that may be produced by smoothness constraints in a section of the mantle where data coverage is poor. Fig. \ref{fig:quiz_2}f features a strong contrast in the character of the seismic structure above and below the 660-km discontinuity. This may be an artefact inherited from the starting model and it may not be possible to remove it by inversions of mantle-transition-zone sensitive waveforms.


Although all of the models include radial anisotropy, the way it is implemented differs significantly. While S40RTS retains the radial anisotropy of its spherically symmetric initial model PREM \citep{Dziewonski_Anderson_1981}, all other models allow the data to introduce lateral variations of anisotropy, but in different forms. REVEAL, for instance, only allows for S wave anisotropy, whereas SEMUCB-WM1 and GLAD-M35 use empirical relations \citep{Montagner_1989} to scale poorly constrained P wave anisotropy to better constrained S wave anisotropy. Similar scaling relations were used to derive 3-D density variations in S40RTS and GLAD-M35, but not in the other models where density was kept equal to the spherically symmetric initial models. These different ways of parameterizing the models control the details of how the data are mapped into 3-D structure. Keeping a poorly constrained parameter like density unchanged or scaling it to some other parameter with an empirical scaling relation has small but noticeable effects on the recovered S velocity structure throughout the whole Earth \citep{Blom_2017}.

All tomographic methods employ approximations to reduce computational cost at the price of, hopefully small, errors in the solution of the forward and inverse problems. Ray theory is an obvious approximation, but there are also less obvious ones. Instead of computing sensitivity kernels with the adjoint method, SEMUCB-WM1 employed an approximation that greatly accelerates convergence towards the optimal model \citep{Li_1995} but may introduce small artefacts \citep{Valentine_2016}. The construction of REVEAL rests on a special type of numerical meshes that reduce computational cost by using less grid points or elements in the direction parallel to the wave front \citep{Thrastarson_2020,Thrastarson_2022}, thereby accepting small numerical errors in the seismic wavefield simulations. The benefit of such wavefield-adapted meshes is the ability to perform hundreds instead of only tens of iterations and to image smaller-scale heterogeneities in the upper mantle that are not present in other FWI models.


\section{Discussion}\label{S:Discussion}

\subsection{Solution to the quiz}\label{SS:resolve_quiz}

Intuitively, we expect slab-like high-velocity anomalies in tomographic cross sections through subduction zones and plume-like low-velocity anomalies beneath hotspots. However, none of the high-velocity structures in Fig. \ref{fig:quiz_1} are within several thousand kilometers of a convergent plate boundary. In contrast, all sections in Fig. \ref{fig:quiz_2}, none of which features a columnar low-velocity anomaly, are centered on a major hotspot (panels a, b on Yellowstone; panel c on Easter; panel d on Iceland; panels e, j on Galapagos; panels f, g, i on Hawaii). The quiz illustrates our innate tendency to be biased by our
expectations. We may see what we want to see, judge the quality of models by the extent to which they satisfy our biases, and create an echo chamber of misinterpretations.

\subsection{Which model is the best?}\label{SS:best_model}

The differences between models and their variable ability to reveal expected features naturally leads to the question of which model is actually the best one. We could wittily fancy $(\text{REVEAL} + \text{S40RTS}) / 2$ the most, but a nuanced answer must address two fundamental issues.

Firstly, the mathematical measures of model quality, such as the resolution matrix or the least-squares misfit introduced in Section \ref{SS:model_quality}, are computed \emph{after} choosing how to parameterize a model, how to discretize the Earth, how to predict traveltimes,  how to compute synthetic waveforms, etc. They quantify variations in data coverage and data quality, but ignore \emph{all the little choices} from Section \ref{SS:choices}. The range of plausible models that are consistent with the data is therefore much larger than implied by misfit measures and resolution matrices. If one would explore the effects of all the little choices, the true quality of a model would always be lower. Unfortunately, it is impossible to exactly quantify how much lower it is.

Secondly, there is no unique metric to quantify quality. Whether a certain quality measure is useful or not depends on the question asked. If one wants to know the average P-wave velocity in the Earth, it is legitimate to divide 12\,742 km by the average traveltime of PKIKP recorded at $180\degree$ epicentral distance. If the PKIKP travel time can be accurately measured, the model is of high quality. Obviously, the same quality measure is not useful if one is interested in small-scale P-wave velocity variations above the core-mantle boundary.

In summary, the straightforward question ``What is the best model?'' is unanswerable even though there is only one Earth and all models are constrained by the same collection of seismograms recorded by the Global Seismic Network. The differences in the maps of Fig. \ref{fig:model_collection} are a natural consequence of letting five research groups construct a seismic model independently with the unavoidable \emph{little choices} to be made. From experience we can guess that the variations in the strength and sharpness of the velocity structures originate from the different applications of damping and smoothing (e.g., Eq. \ref{E:methods012}) and the change in the velocity patterns across the 660-km discontinuity (see Fig. \ref{fig:quiz_1} and \ref{fig:quiz_2}) is caused by a split parameterization of the upper and lower mantle.

We are admittedly evasive, but we can offer the pragmatic advice to engage the producer of a model in the science questions and the model interpretation. Although uncertainties are hard to quantify, tomographers often have good intuition of the actual reliability of their models and the origin of the anomalies. For example, the upper mantle is best constrained by surface waves (Fig. \ref{fig:phasevel_maps}) and D'' by diffracted body waves (Fig. \ref{fig:sdiff}). How anomalies appear may depend on the size of the surface-waves and diffracted S-wave data sets, how the uppermost and lowermost are discretized, and whether seismic anisotropy is included in the parameterization. It is fair to analyze multiple models and to consider differences as a first-order estimate of the quality of the model. It is questionable to select solely the most popular model, the most accessible model, or one's own.

\subsection{Community Monte Carlo}

Differences in tomographic models may raise the question whether seismic tomography is actually useful. However, model differences must exist because model uncertainties due to (i) data coverage and quality and (ii) the range of justifiable technical choices exist. The uncertainty of the model is not an issue; quantifying it is the true challenge.

From this perspective, the diversity of tomographic models is an opportunity to map out the true uncertainties. As different research groups choose different data collections and methodologies, they sample the actual range of plausible models. This is similar and complementary to the application of Monte Carlo methods that systematically generate random (velocity) models that explain the data to within their errors \citep[e.g.,][]{Mosegaard_1995,SambridgeMosegaard_2002}, while keeping the data collections and methodologies constant. In this sense, research groups that produce different tomographic models perform \emph{Community Monte Carlo} sampling that explores the allowable range of subjective choices. Just as in regular Monte Carlo sampling, Community Monte Carlo can map out the true uncertainties only when the samples are sufficiently diverse. We should therefore make an effort to produce more different tomographic models and avoid aiming for similar ones.

That the ensemble of little choices significantly affects inversion results, and interpretations and decisions based upon them, is not a particularity of seismic tomography. It is a general property of realistic inverse problems where data are imperfect and incomplete, and modeling techniques must be simplified to become computationally tractable. Other examples include ice sheet, sea level and climate projections, where Community Monte Carlo is already being practiced to some extent \citep[e.g.,][]{Eyring_2016,edwards_etal_2021_nature,seroussi_etal_2024_ef}.

\subsection{Propagating seismic model uncertainties into geodynamic models}

Measures of quality or uncertainty in tomographic models not only ignore the effect of subjective choices, they are also difficult to propagate quantitatively into subsequent interpretations. For example, how shall we include a tomographic point-spread function or resolution length into a geodynamic data assimilation method that aims to infer past mantle flow? This raises the question if the quality measures that we can compute in seismic tomography with reasonable effort are actually useful for the users. Just as the notion of "best", also the notion of "useful" depends on the problem that one wishes to solve. Does this mean that seismic tomographers should provide a quality measure for all imaginable uses of their models? 

The only viable solution seems to be the generation of an ensemble of plausible tomographic Earth models. This should ideally involve both Community Monte Carlo sampling (probing the space of subjective choices) and regular Monte Carlo sampling (for a fixed set of choices). With currently available computational resources, Monte Carlo sampling is still challenging, but approximate ensemble generators such as nullspace shuttles are feasible already today \citep[e.g.,][]{Deal_1996,Fichtner_2019,Keating_2024}. Performing, for example, a geodynamic data assimilation for each ensemble member, will provide an ensemble of plausible mantle flow models that represents the uncertainties in this specific inference.

\subsection{Outlook}

Increasing the quantity and quality of seismic data has always been and will continue to be the single most important contributor to improved tomographic images of the Earth. The distribution of global seismic networks today is not different than twenty years ago. Major data gaps remain to be filled, for example, in the oceans, polar regions, and in countries where station coverage is sparse or data are not publicly available. Promising developments include large deployments of ocean-bottom seismometers \citep[e.g.,][]{Laske_2009,Staehler_2016,NIED_2019}, seismometer floats that traverse the oceans autonomously \citep[e.g.,][]{Simons_2009,Joubert_2016}, and emerging technologies that use existing telecommunication fibers for earthquake sensing \citep[e.g.,][]{Marra_2018,Marra_2022,Bogris_2022,Noe_2023}.

Novel data need to be complemented by faster computers and more efficient wave simulation methods that allow us to actually use them to improve tomographic images. Current global-scale FWI \citep[e.g.,][]{French_2014,Cui_2024,Thrastarson_2024}, for example, operates at minimum periods of several tens of seconds, which means that valuable travel time data of short-period (around 1 s) scattered P waves \citep[e.g.][]{hedlin_1997, kaneshima_helffrich_1998, Rost_2003} cannot be included. In contrast, tomography based on ray theory \citep[e.g.,][]{Ritsema_2011,Simmons_2021}, can exploit high-frequency body waves but fails to produce accurate synthetic seismograms for the strongly heterogeneous parts of the Earth, such as the lithosphere. The grand unifying method of seismic tomography remains to be developed.

Finally, we remark that meaningful interpretations of seismic tomography depend on collaborations. Tomographic images have long been the most important products for studies of deep-earth structure dynamics. It takes the expertise and the candor of the producers to explain how the reds and blues are constrained by data and how ``all the little choices'' mask modeling uncertainties.

\section{Glossary}\label{S:Glossary}

As any other field, seismic tomography is infiltrated by jargon that is difficult to understand, may have unexpected meaning, or sometimes does not actually carry any meaning at all. Below we provide a short glossary of jargon that we have not yet defined in the previous sections.\\[5pt]
\textbf{Adjoint tomography} is one of numerous FWI variants that have been named to reflect methodological details. Although the tomography of which adjoint tomography is actually the adjoint does not exist, this term is often used for FWI flavours based on misfit functions that measure travel time differences and employ the adjoint method to compute sensitivity kernels.\\[5pt]
\textbf{Banana-doughnut kernel} is a culinary reference first made by \citet{Marquering_1999} to a special type of 3-D sensitivity kernel, similar to the one shown for the 2-D case in Fig. \ref{fig:finite_frequency}b. These kernels resemble a doughnut when cut perpendicular to the ray path and a banana when cut parallel to the ray path.\\[5pt]
\textbf{Checkerboard tests} use artificial input Earth models, often resembling a checkerboard, to compute artificial data that replace real data in a seismic tomography. The difference between the input and the tomographic model serves as a proxy for the achievable resolution. Although methodologically simple, such recovery tests can be misleading because the choice of input structure is subjective, and realistic data noise and forward modeling errors are typically ignored. Hence, producing a good-looking checkerboard test is generally much easier than producing a meaningful tomographic model based on real data.\\[5pt]
\textbf{Classical tomography} is elegant and timeless, like classical Greek architecture or the classical music of Mozart. Yet, the term is being abused as a synonym of old-fashioned, ignoring the fact that simple and efficient methods that have been developed earlier (e.g., ray-based travel time tomography) are indeed more elegant than later additions to the tomographic toolbox that require heavy machinery and complicated workflows (e.g., FWI).\\[5pt]
\textbf{High-resolution} is an adjective frequently used to emphasize the merits of a particular method or data set, but its actual meaning is unclear. How high exactly, and what is a low-resolution model? We suggest to not use this term unless a precise definition can be provided.\\[5pt]
\textbf{Imaging} is a term with at least two different meanings. On the one hand, it is a synonym for migration (see below), i.e., methods to infer the location of reflectors. On the other hand, it is a catch-all phrase for any method that turns data into an image of the Earth's interior that can be displayed on a screen and somehow interpreted. In the latter sense, tomography is also an imaging method.\\[5pt] 
\textbf{Migration} is a term primarily used in seismic exploration. It loosely encompasses all methods that use seismic data to infer the location of reflectors, i.e., sharp material contrasts, at depth. A widely used migration method is \emph{reverse-time migration} \citep[e.g.,][]{Baysal_1983}, whereby time-reversed recordings of reflected waves are used as sources in numerical simulations. The resulting numerical wavefield tends to focus near the locations of reflectors in the subsurface, which produces a reflector image.\\[5pt]
\textbf{Normal modes} are whole-Earth oscillations that are excited by large earthquakes. Similar to the normal modes of guitar strings, the frequencies (or the 'sound') of the Earth's normal modes are controlled by its mechanical properties (e.g., length of string, tension, density). Normal-mode observations can be used to constrain the Earth's large-scale ($> 5000$ km) structure, and perhaps the 3-D density variations in the lower mantle \citep[e.g.,][]{Koelemeijer_2017, Lau_2017}.\\[5pt]
\textbf{Reference models} serve as a reference, e.g., for 3-D velocity variations \citep[PREM,][]{Dziewonski_Anderson_1981} or body wave travel times \citep[ak135,][]{Kennett_1995} relative to those of an average 1-D Earth model. The term is sometimes used incorrectly to claim outstanding importance or quality of some model (similar to the use of \emph{high-resolution}) without specifying for what it is supposed to be a reference.\\[5pt]
\textbf{Slabs and plumes}: An eminent colleague once quipped that slabs are high-velocity anomalies traversed by a bundle of rays and that plumes are low-velocity anomalies crossed by two or three rays. The remark is a wink to readers eager to perform well on the quiz of section \ref{S:Quiz}.


\begin{acknowledgments}
We, and certainly many colleagues, gratefully acknowledge the work of seismic network operators who openly share their data. Progress in seismic tomography would not be possible without. This work was supported by the Swiss National Supercomputing Center (CSCS), ETH Z\"{u}rich and the Great Lakes HPC Cluster at the University of Michigan. Jeroen Ritsema acknowledges support from the United States National Science Foundation via award EAR--2019379. Constructive comments by Hans-Peter Bunge, Mohammed Almarzoug, Thomas Hudson, Noami Kaplunov, Carl Schiller and Evan Delaney helped us to significantly improve the readability of this paper.
\end{acknowledgments}


\bibliography{biblio}
 
\end{document}